\documentclass[final,3p,times,twocolumn]{elsarticle}

\usepackage{amssymb}
\usepackage{amsmath}

\usepackage{color}
\usepackage{soul}
\usepackage{cases}

\definecolor{vs}{rgb}{0.1,0.4,0.1}                  
\usepackage{hyperref}
\usepackage{url}
\usepackage{flushend}

\journal{Physica D}

\begin{document}

\begin{frontmatter}



\title{The Eckhaus instability: from initial to final stages}


\author{Michael I. Tribelsky} 

\affiliation{organization={Faculty of Physics, Lomonosov Moscow State University},
            addressline={Leninskie Gory, 1, Bldg. 2}, 
            city={Moscow},
            postcode={119991}, 
            country={Russia}}

\begin{abstract}
A systematic analysis of the Eckhaus instability in the one-dimensional Ginzburg-Landau equation is presented. The analysis is based on numerical integration of the equation in a large (xt)-domain. The initial conditions correspond to a stationary, unstable spatially periodic solution perturbed by "noise." The latter consists of a set of spatially periodic modes with small amplitudes and random phases. The evolution of the solution is examined by analyzing and comparing the dynamics of three key characteristics: the solution itself, its spatial spectrum, and the value of the Lyapunov functional. All calculations exhibit four distinct, mutually agreed, well-defined regimes: (i) rapid decay of stable perturbations;  (ii) latent changes, when the solution and the Lyapunov functional undergo minimal alterations while the Fourier spectrum concentrates around the most unstable perturbations;  (iii) a phase-slip period, characterized by a sharp decrease in the Lyapunov functional; (iv) slow relaxation to a final stable state.  
\end{abstract}
%
%


\end{frontmatter}



\section{Introduction}
\label{sec:Introd}

The Ginzburg-Landau (GL) equation is the most well-known equation in solid-state physics, nonlinear physics, and beyond. The Nobel Prize awarded to its creators is strong evidence of its significance. Initially introduced in the phenomenological theory of superconductivity, the GL equation has proven to be extremely important in many other fields, particularly in theory of pattern formation in nonequilibrium systems~\cite{Cross1993}. It is explained by the fact that in many cases, the initial underlying equations describing pattern formation close to onset may be reduced to the GL equation, under very general assumptions~\cite{Tribelskii1997}.

In the one-dimensional case, the appropriate scaling of the variables reduces the equation to the following parameter-free universal form:

\begin{equation}\label{eq:GL_parameter-free}
  \frac{\partial\psi}{\partial t} = \frac{\partial^2\psi}{\partial x^2}+\left(1 -|\psi|^2\right)\psi,
\end{equation}
where $t$, $r$, and $\psi$ stand for time, spatial coordinate and complex order parameter, respectively; and all these quantities in Eq.~\eqref{eq:GL_parameter-free} are dimensionless.

Needless to mention numerous publications devoted to this equation and its solutions --- they are well-known to any expert in the field. After many years of study, anything related to this equation may seem already understood. However, an important aspect of the evolution of its unstable stationary solutions remains unexplored. The goal of this study is to address this issue. To clarify the problem in question further, it is important to recall some known features of the GL equation; see, e.g.,~\cite{Cross1993}.

The equation~\eqref{eq:GL_parameter-free} has a trivial solution given by $\psi = 0$. Its linear stability analysis against perturbations of the form $\delta\psi = \text{const} \cdot e^{\gamma t + ikx}$ yields the spectrum.
\begin{equation}\label{eq:gammaGL}
  \gamma = 1 - k^2,
\end{equation}
i.e., the solutions is unstable against perturbations with $|k|<1$. 

The equation has other stationary solutions of the form 
\begin{equation}\label{eq:psiGL}
  \psi = R{\rm e}^{i\varphi};\; R=\sqrt{1 - k^2};\;\varphi = kx+const;\; -1\leq k \leq 1.
\end{equation}
In the case $-\infty <x<\infty$, the constant in Eq.~\eqref{eq:psiGL} can be set to zero by the corresponding shift of the origin of the $x$-axis. Therefore, we can drop it without loss of generality.

Consider now the GL equation in a finite domain $-L/2 \leq x \leq L/2$ with periodic boundary conditions at $x=\pm L/2$ and introduce the quantity ({\it the Lyapunov functional density}) 
\begin{equation}\label{eq:Lyapunov_GL}
\mathcal{F} = \frac{1}{L}\int_{L/2}^{L/2}\left(\left|\frac{\partial\psi}{\partial x}\right|^2 -|\psi|^2+\frac{1}{2}|\psi|^4\right)dx.
\end{equation}  
Then, Eq.~\eqref{eq:GL_parameter-free} may be presented as 
\begin{equation}\label{eq:gradient_eqs}
 \frac{\partial \psi}{\partial t}=-\frac{\delta \mathcal{F}}{\delta \psi^\ast},
\end{equation}
where ${\delta \mathcal{F}}/{\delta \psi^\ast}$ stands for the variational derivative, and $\psi^\ast$ designates the complex conjugation. 

Note that periodic boundary conditions impose a weak conservation law:

\begin{equation}\label{eq:Phi}
  \Phi(t) \equiv \text{Arg}[\psi(L/2, t)] - \text{Arg}[\psi(-L/2, t)] = 2\pi n,
\end{equation}
where Arg represents the phase of a complex quantity, and $ n $ is an integer. It indicates that $ \Phi(t) $ can change only in discrete steps, specifically integer multiples of $ 2\pi $. In between these steps, $ \Phi(t) $ remains constant. These changes correspond to the so-called {\rm phase slip} processes, which occur when $ \psi $ vanishes at some values of $x= x_{\rm ps} $ and $t= t_ {\rm ps}$, so that the phase of $\psi(x_{\rm ps},t_{\rm ps})$ becomes uncertain~\cite{langer1967intrinsic, ivlev1984theory, Tribelsky1992, Yamazaky1992, Tribelsky1995}. We will revisit this topic later.     

A remarkable feature of Eqs.~\eqref{eq:Lyapunov_GL},~\eqref{eq:gradient_eqs} is that
\begin{equation}\label{eq:dLyap/dt}
  \frac{d\mathcal{F}}{dt} = -\frac{2}{L}\int_{-L/2}^{L/2}\left|\frac{\partial\psi}{\partial t}\right|^2dx \leq 0.
\end{equation}
That is to say, any dynamics decrease the Lyapunov functional. Limiting transition $L \to \infty$ extends these results to the case $-\infty < x < \infty$. 

Linear stability analysis of solution \eqref{eq:psiGL} against perturbations $\delta R \propto \delta \varphi \propto \exp(\sigma t + iqx)$ yields the spectrum

\begin{equation}\label{eq:GL_sigma(k,p)}
  \sigma(k,q) = k^2-1-q^2 + \sqrt{(1-k^2)^2+4 k^2q^2},
\end{equation} 
The smallness of \(\delta\varphi\) allows us to express \mbox{\(\exp[i(kx + c\exp(\sigma t + iqx))]\),} where \(c\) is an infinitesimal constant, as \mbox{\(\exp(ikx)[1 + c\exp(\sigma t + iqx)]\).} In other words, for the given ansatz, the actual wavenumber of the perturbation is \(k + q\).

A straightforward inspection of Eq.~\eqref{eq:GL_sigma(k,p)} leads to the Eckhaus stability criterion: the solution \eqref{eq:psiGL} is stable, provided \(|k| \leq k_{E}\), and is unstable otherwise~\cite{Eckhaus1965}. Here, \(k_{E} = 1/\sqrt{3} \approx 0.577\).

The results indicate that an initially spatially periodic pattern with \(k > k_{E}\) is unstable, as it succumbs to the growth of unstable modes with \(\sigma(k, q) > 0\). Conversely, the noted monotonic decrease of the Lyapunov functional over time suggests that the system's phase trajectory in the corresponding functional space must either converge to a fixed point (a stable stationary state) or go ``to infinity'', resulting in \(\mathcal{F} \to -\infty\) as \(t \to \infty\). However, the latter scenario cannot occur. Indeed, the only negative contribution to the integrand in Eq.~\eqref{eq:Lyapunov_GL} is related to \(-|\psi|^2\). Therefore, an unlimited decrease of \(\mathcal{F}\) would only be associated with this term as \(|\psi| \to \infty\). Yet, at large \(|\psi|\), the contribution of \(-|\psi|^2\) becomes less significant than that of the positive term \(|\psi|^4/2\). Thus, as \(|\psi| \to \infty\), the Lyapunov functional actually would increase, which contradicts Eq.~\eqref{eq:dLyap/dt}.

In conclusion, the dynamics resulting from the decomposition of an unstable stationary solution must end with the emergence of another stationary solution, which is stable. Considering that the only set of stable stationary solutions of the GL equation is described by Eq.~\eqref{eq:psiGL}\footnote{The GL equation also has another set of stationary solutions expressed in terms of elliptic integrals, which correspond to saddle points in the functional space; all these solutions are unstable~\cite{kramer1985eckhaus, tribelsky1991transitions}.}, we conclude that the dynamics must result in the formation of a new solution of the type described by the sme Eq.~\eqref{eq:psiGL} but with \(|k| \leq k_{E}\). In other words, these dynamics represent a problem known as {\it wave number selection in pattern formation}~\cite{Cross1993}.

If we now examine Eq~\eqref{eq:GL_sigma(k,p)}, which presents the spectrum of the linear stability analysis of the solutions in question, we find that $\sigma$ is an even function of $q$. It indicates that pairs of unstable modes with the compound wavenumbers $k \pm q$ share the same growth rate. On the other hand, the above reasoning suggests that the final state of the instability evolution must result in a spatially periodic pattern characterized by a {\it single\/} wavenumber. 

Thus, the linear stage of the Eckhaus instability cannot capture the entire process, even qualitatively; a comprehensive nonlinear approach is required. This fact is well-established. Since Wiktor Eckhaus's pioneering work in 1965~\cite{Eckhaus1965}, research on Eckhaus instability has continued to evolve up to now~\cite {gianfrani2025eckhaus,Tsubota2024,Xiang2025}. Numerous publications (e.g.,  ~\cite{kramer1985eckhaus,kramer1988pattern,boucif1984role,lowe1985pattern,dominguez1986eckhaus,pocheau1987convective,simon1988solitary,rashkeev1999irregularities}) address various aspects of linear and nonlinear stages of Eckhaus instability. 

However, despite this issue's extensive literature and long history, a complete description of the nonlinear stages of Eckhaus instability has yet to be achieved. What follows is an attempt to provide this description. 

\section{Problem formulation}
\label{sec:Problem}
We tackle the problem through the numerical integration of the GL equation. The phase of the wave function \( \psi \) becomes singular at the phase slip points, while the real and imaginary parts of \( \psi \) remain there regular functions of \( x \) and \( t \). As a result, a standard method for numerical integration in this context is to express \( \psi \) using its real and imaginary components, rather than its modulus and phase; see, for example, ~\cite{Tribelsky1992}.

The simulations are conducted over a large \( x \) domain defined by \( l/2 \leq x \leq L/2 \), with periodic boundary conditions applied at \( x = \pm L/2 \). We also specify the following initial condition:
\begin{equation}\label{eq:psi0_num}
  \psi_0(x) = \sqrt{1-k_0^2}\,{\rm e}^{ik_0x}+A\sum_{n=-N}^{N}e^{i(n\Delta kx +\varphi_{0n})},
\end{equation}
where $k_0>k_{E}$, $\Delta k = 2\pi/L \ll k_0$, $N \gg 1$, $A$ is a real constant much smaller $\sqrt{1-k_0^2}$, and $\varphi_{0n}$ are random initial phases ($-\pi \leq \varphi_{0n} \leq \pi$). The calculations indicate that if these conditions hold, the dynamics always exhibit the same behavior weakly dependent on the specific choice of $k_0,\;L,\;N$ and $\varphi_{0n}$. It makes possible to extend the results obtained to the case $-\infty <x<\infty$. 

\section{Results and discussion}

In this section, we present the results of a typical simulation at the following values of the problem parameters: $k_0 = 0.8;\;\sqrt{1-k_0^2}=0.6;$ \mbox{$A= 0.005;\;N=48;\;\Delta k = 0.05$.} This $\Delta k$ corresponds to $L\approx 125.664$. 

For reference, it is also worth presenting the values of several characteristic quantities of the problem at the specified values of the parameters. In particular, at the given value of $N$, the maximal compound wave number of the perturbations $k_0+q_{_N} = \Delta kN = 2.4$, i.e., $q_{_N} =1.6$. According to Eq.~\eqref{eq:GL_sigma(k,p)}, it results in $\sigma(q_{_N},k_0)\approx-0.3348$. That is to say, the last taken into account mode in Eq.~\eqref{eq:psi0_num} lies already rather far in the stable region. According to Eq.~\eqref{eq:GL_sigma(k,p)} the value of $q$ maximizing $\sigma(k_0,q)$ equals $q_{\rm max} \approx 0.7677$. In other words, according to the linear stability analysis, the most unstable perturbations have the compound wavenumbers $k_0 \pm q_{\rm max}$ equal to 0.0323 and 1.5677, respectively. The corresponding growth rate $\sigma(k_0,q_{\rm max}) \approx 0.3306$. 

\begin{figure}
  \centering
  \includegraphics[width=.8\columnwidth]{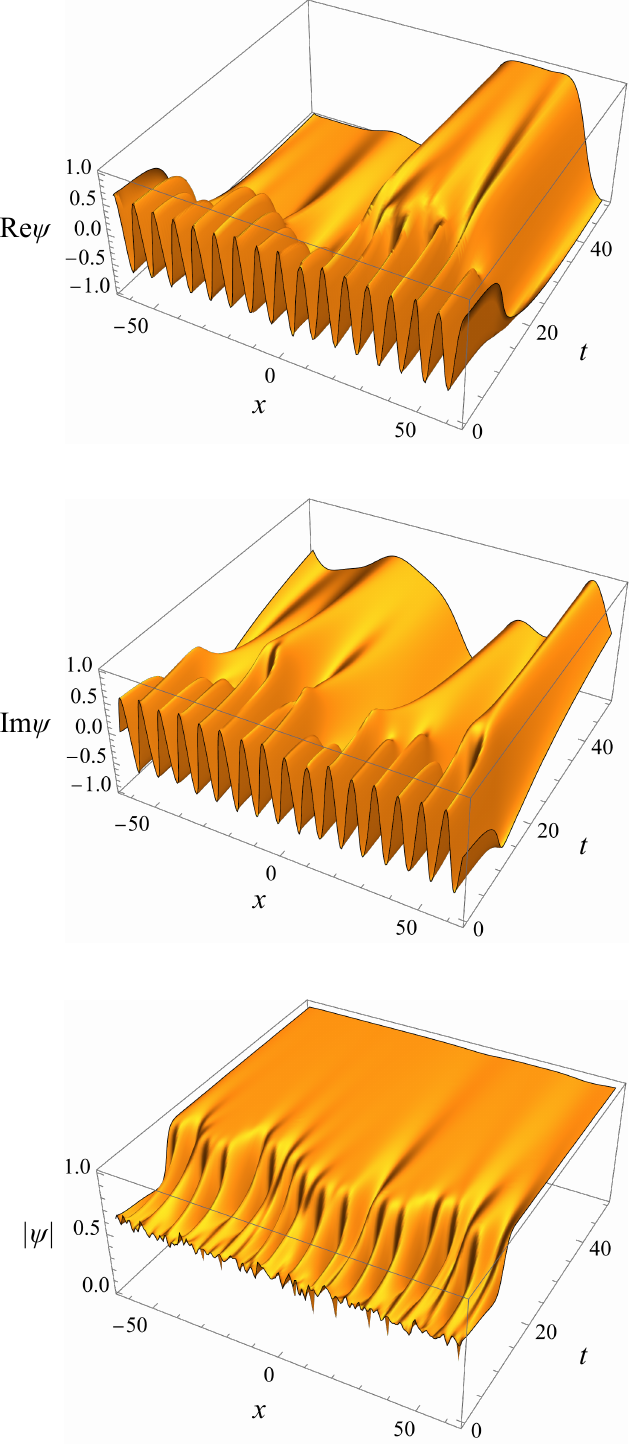}
  \caption{Profiles of Re$\,\psi(x,t)$, Im$\,\psi(x,t)$, and $|\psi(x,t)|$.}\label{fig:psi}
\end{figure}

Figure \ref{fig:psi} illustrates the spatiotemporal evolution of $\psi(x,t)$ for the time interval $0 \leq t \leq 50$. Initially, there are no significant changes until $t$ reaches approximately~10. After this moment, $\psi(x,t)$ undergoes drastic transformations. By $t \approx 20$, the changes in the modulus of $\psi$ have mostly been completed, while the alterations in its real and imaginary parts persist; however, their evolutions become much slower. 

\begin{figure}
  \includegraphics[width=\columnwidth]{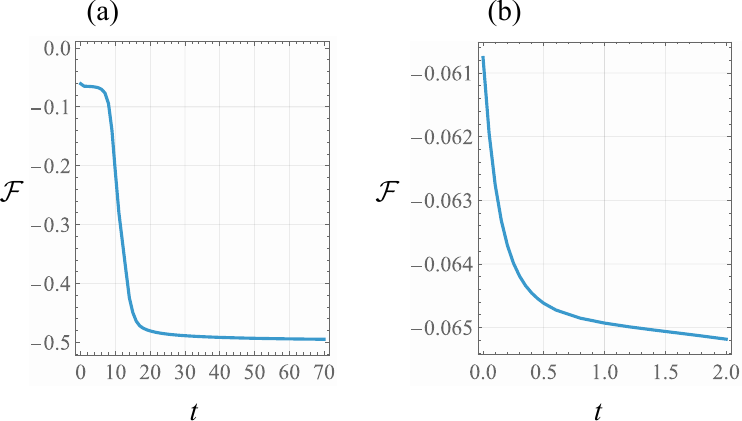}
  \caption{(a) Temporal evolution at $0\leq t \leq 70$ of the Lyapunov functional for the GL equation with the initial conditions given by Eq.~\eqref{eq:psi0_num}. (b) Zoom of the initial part of this profile.}\label{fig:FGL}
\end{figure}

Figure \ref{fig:FGL} depicts the corresponding Lyapunov functional dynamic. Four distinct regimes (or stages) can be observed: 

\begin{description}

  \item[(i)] A sharp decrease in $\mathcal{F}(t)$ from its initial value \mbox{$\mathcal{F}(0)\approx -0.0608$} to about -0,0652 at $t=2$; see Fig.~\ref{fig:FGL}(b). 
  \item[(ii)] A very slow variation at $2\leq t \leq 8$. 
  \item[(iii)] Another sharp decrease at $8\leq t \leq 15$.
  \item[(iv)] A slow asymptotical transient to a stationary state at $t>15$.

\end{description} 

\begin{figure}
  \includegraphics[width=\columnwidth]{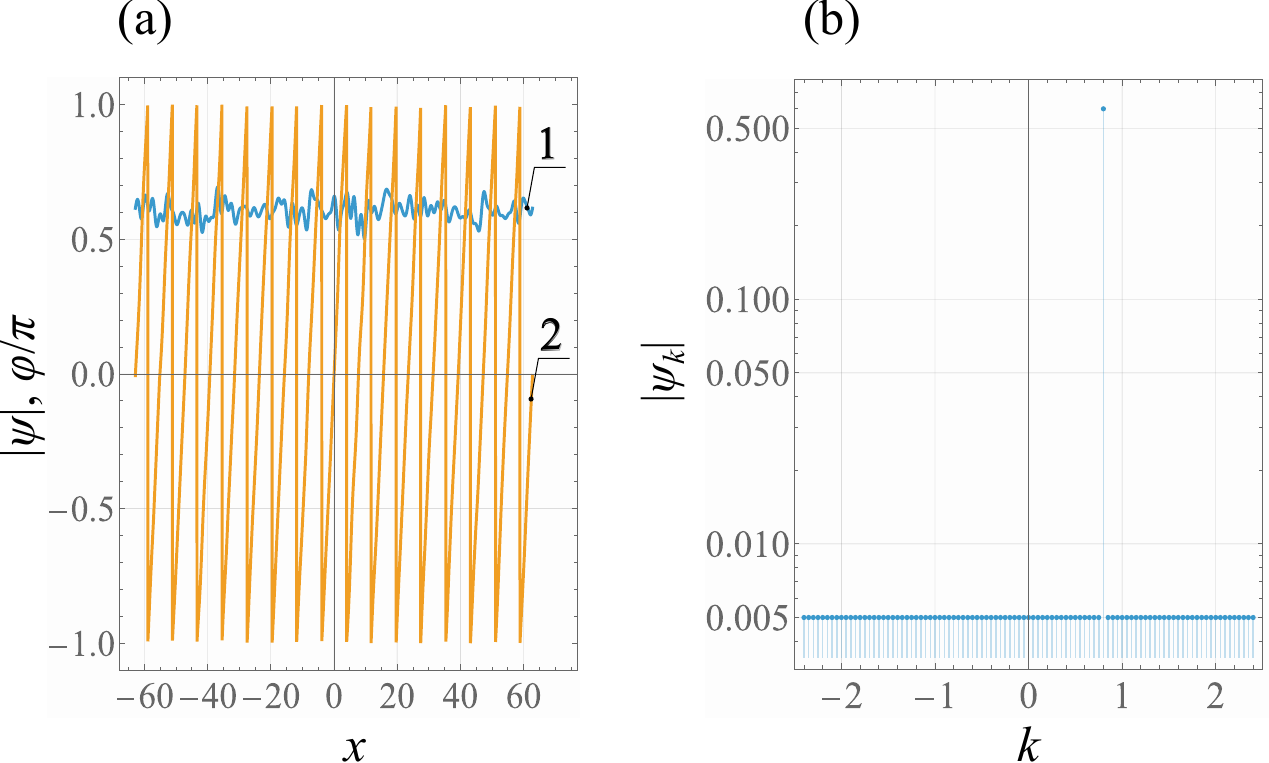}
  \caption{(a) The initial profile of the modulus of \( \psi \) at \( t=0 \) (marked by number 1) is shown, along with its phase normalized by \( \pi \) (marked by number 2). The phase, denoted as \( \varphi \), is confined to the range \mbox{\( -\pi \leq \varphi \leq \pi \).} If the continuous phase value exceeds this domain, it is manually adjusted by adding (or subtracting) the appropriate integer multiples of \( 2\pi \). This adjustment introduces discontinuities in the corresponding profile. (b) A log plot displays the moduli of the coefficients from the Fourier series expansion of \( \psi(x,0) \).}\label{fig:t0}
\end{figure}

\begin{figure}[h!]
  \includegraphics[width=\columnwidth]{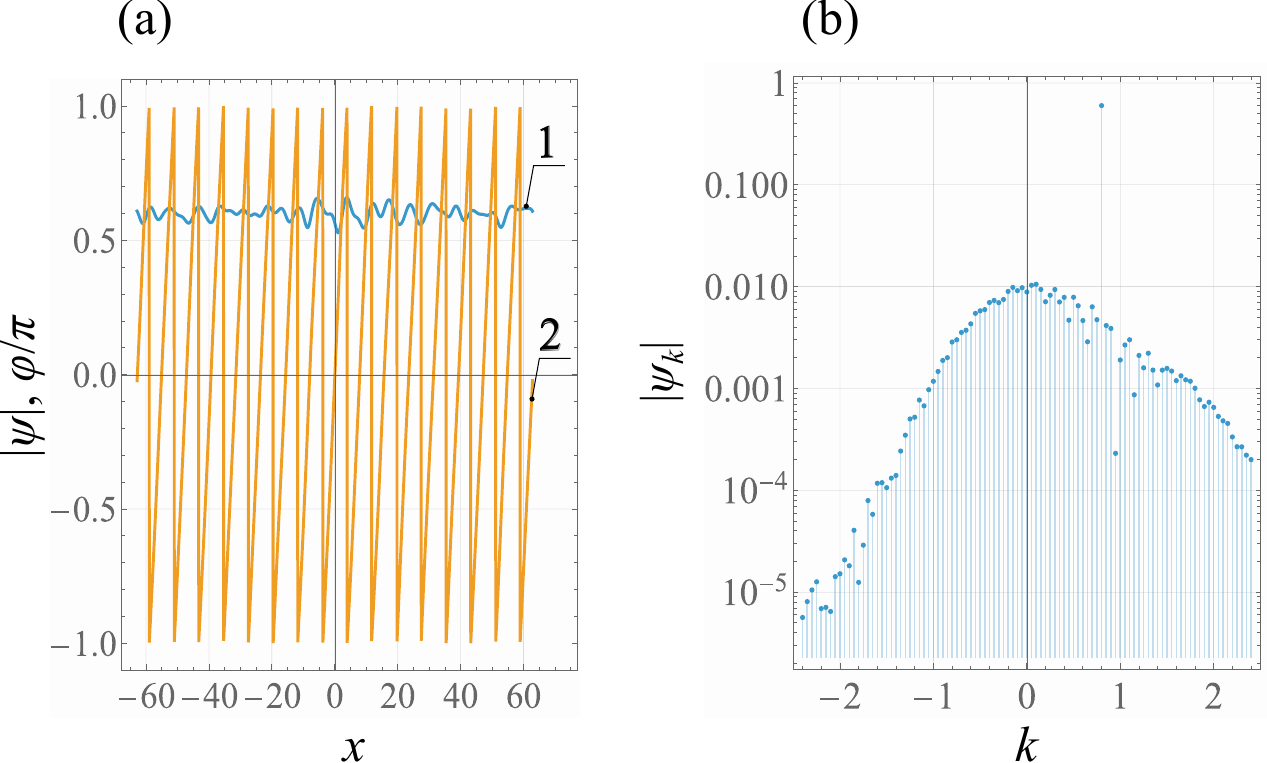}
  \caption{The same profiles as those in Fig.~\ref{fig:t0} at $t=2$.}\label{fig:t2}
\end{figure}

\begin{figure}[h!]
  \vspace*{-3pt}
  \includegraphics[width=\columnwidth]{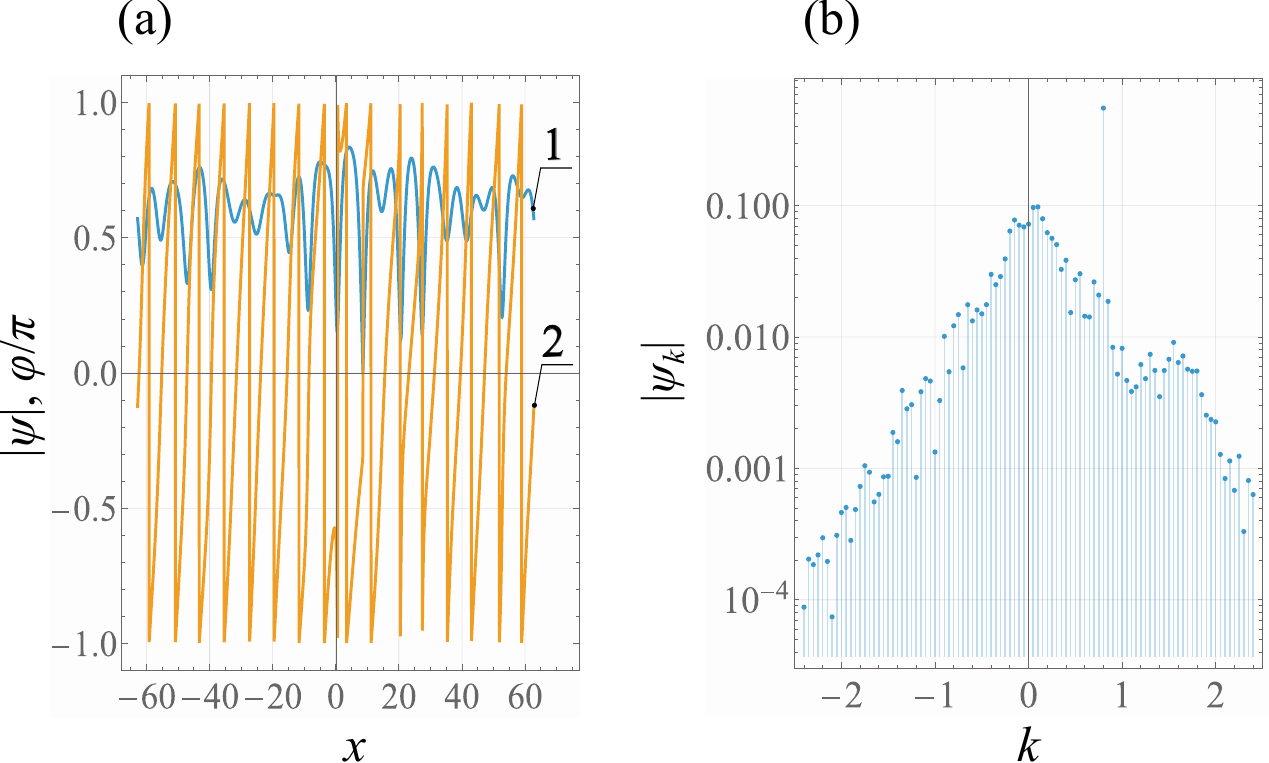}
  \caption{The same profiles as those in Fig.~\ref{fig:t0} at $t=8$.}\label{fig:t8}
\end{figure}

\begin{figure}
  \includegraphics[width=.96\columnwidth]{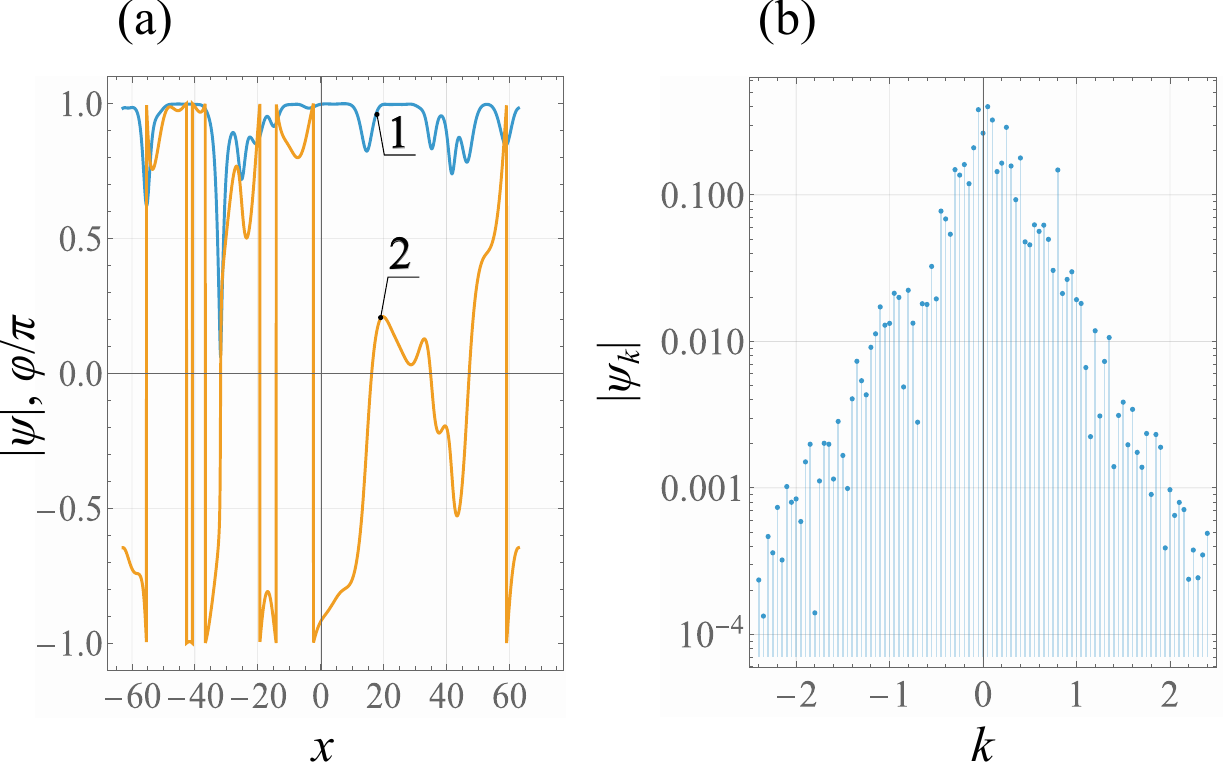}
  \caption{The same profiles as those in Fig.~\ref{fig:t0} at $t=14$.}\label{fig:t14}
\end{figure}

\begin{figure}[h!]
  \includegraphics[width=\columnwidth]{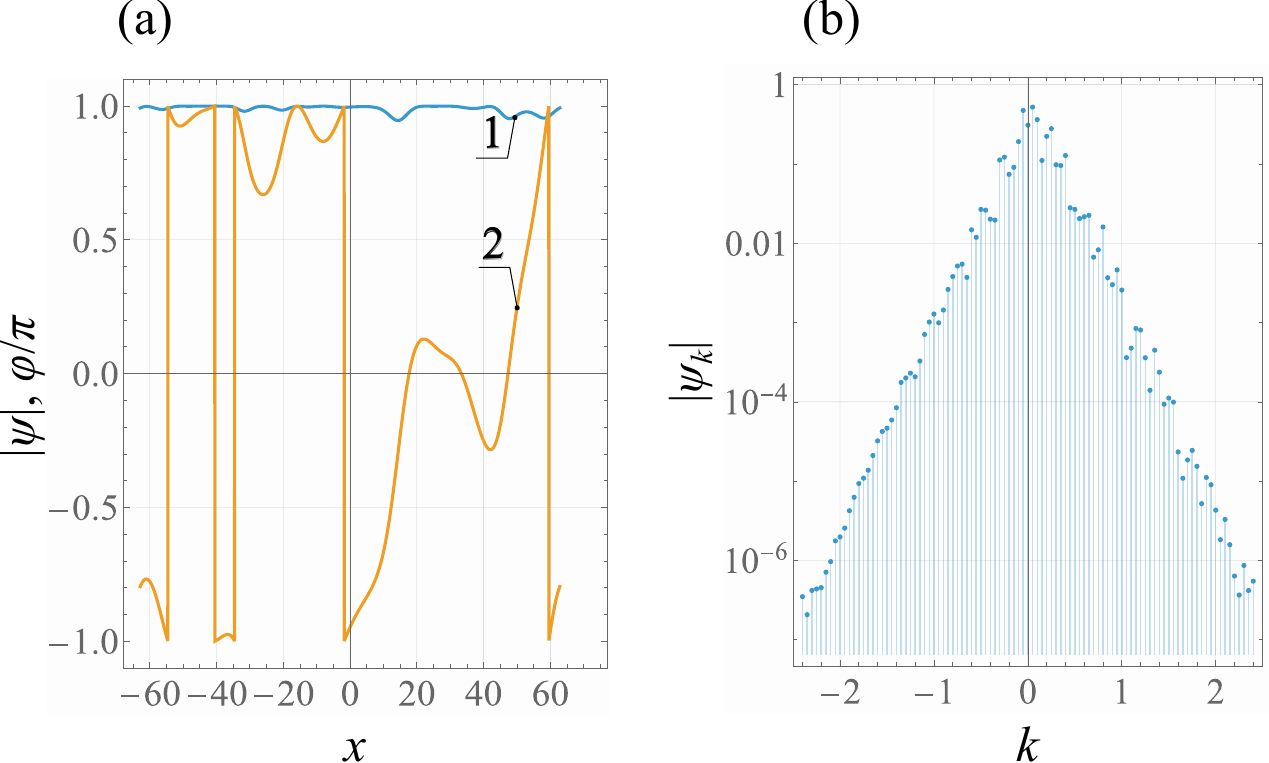}
  \caption{The same profiles as those in Fig.~\ref{fig:t0} at $t=20$.}\label{fig:t20}
\end{figure}

To understand this behavior we inspect the profiles $|\psi(x,t)|$, $\varphi(x,t)\equiv {\rm Arg}[\psi(x,t)]$, and the moduli of the corresponding coefficients $|\psi_k(t)|$ of the expansion of $\psi(x,t)$ in the Fourier series:
\begin{equation}\label{eq:psi_Fourier_ser}
  \psi(x,t) = \sum_k \psi_k(t) {\rm e}^{ikx};\;\; \psi_k(t) =\frac{1}{L}\int_{-L/2}^{L/2}\psi(x,t)e^{-ikx}dx,
\end{equation}
where $k=n\Delta k$, and $n$ is an integer. Let us examine the variations of these profiles at each of the specified stages; see Figs~\ref{fig:t0} to \ref{fig:t20}. 

{\bf (i)} At $0\leq t \leq 2$, the stable modes of the initial perturbation are rapidly suppressed, which explains the sharp change in the Lyapunov functional during this period. By time \(t=2\), all of these modes are mostly died off. However, since their initial amplitudes are small, this rapid suppression leads to only minimal changes in the value of \(\mathcal{F}\); see Fig.~\ref{fig:FGL}.

In this stage, the amplitude of the initial unstable mode with $k=k_0$ remains substantially larger the amplitudes of all other modes. For this reason, the phase of \(\psi\) is not significantly affected, including both its profile and the phase difference $\Phi$ between the edges of the segment --- the latter remains constant 
equal its initial value, \(\Phi_0 = Lk_0\). According to our choice of the problem constants, it is $2\pi k_0 / \Delta k = 2\pi\cdot 16$. Regarding \(|\psi|\), while the mean amplitude of its modulations remains nearly constant, the removal of high harmonics from its profile results in a smoother appearance. This is illustrated in Figs. \ref{fig:t0}a and \ref{fig:t2}a.

\begin{figure}
  \includegraphics[width=.8\columnwidth]{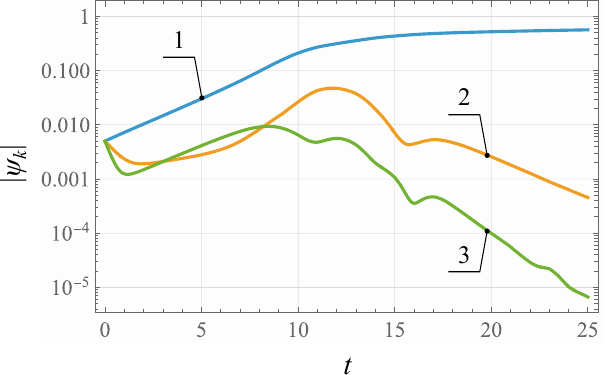}
  \caption{Profiles $|\psi_k(t)|$. (1) $k=0.05$; (2) $k=1$; (3) $k=1.55$.}\label{fig:psi_0.1_1_1.55}
\end{figure}

{\bf (ii)} Latent changes characterize this period. During this time, the amplitude of the primary initial mode with \( k = k_0 \) gradually decreases but still remains much larger the amplitudes of all other modes. Regarding them, the most unstable modes, according to the linear stability analysis (which, in our case, have wavenumbers close to \( k = \Delta k \)), grow nearly exponentially, with the growth rates approximately equal to that given by Eq.~\eqref{eq:GL_sigma(k,p)}. Other unstable modes also experience growth, but it occurs with slower rates and even can be non-monotonic; see below the discussion of Fig.~\ref{fig:psi_0.1_1_1.55}. 

As a result, the spectrum \( |\psi_k| \) becomes sharper. This gradual sharpening of the spectrum, along with its asymptotic convergence to the wavenumber corresponding to the final stable state --- resulting from the Eckhaus instability --- is a global feature of this phenomenon, as seen in the spectra presented in Figs.~\ref{fig:t0} to~\ref{fig:t20}.

For the stage in question, these dynamics have a minimal effect on the Lyapunov functional. The phase difference \( \Phi \) also remains at its initial value. The stage of latent changes continues until the amplitudes of the most unstable modes become comparable to the amplitude of the mode with \mbox{\( k = k_0 \).} After this, stage (iii) begins.

{\bf (iii)} During this stage, the superposition of modes with different values of \( k \) can cause the magnitude of \( |\psi| \) to vanish at certain  \( x \) and \( t \). As we already stressed, this phenomenon leads to phase slips. In fact, calculations indicate that numerous phase slips occur during this stage. Each phase slip decreases \( \Phi \) by \( 2\pi \), allowing \( \psi \) to eliminate the extra phase~\footnote{See Ref.~\cite{Tribelsky1995} for a comprehensive geometrical explanation of the phase slips. It also explains why, in the general case, a phase slip gives rise to the $2\pi$ phase shift (not to a lager integer multiple of $2\pi$).}.


Figures \ref{fig:t8} and \ref{fig:t14} illustrate the beginning and end of this stage. Figure \ref{fig:t8} corresponds to \(t = 8\). Just before this moment, a phase slip occurs at \(x \approx 0.487\) and \(t \approx 7.606\). At this point, we still observe a sharp minimum in \(|\psi|\) at \(x \approx 0.487\), but this minimum is going to disappear, leading to an increase in the modulus of \(\psi\) at this point for \(t > 8\).

Additionally, there are many phase slips to occur along the way, each corresponding to sharp local minima of \(|\psi|\). At these minima, the modulus of \(\psi\) decreases for \(t > 8\) until, at certain moments, \(|\psi|\) vanishes, resulting in a phase slip. After this, \(|\psi|\) relaxes to its mean value.

The phase slip generation stage has a well-defined time frame. It begins with the occurrence of the first phase slip at \(t \approx 7.606\) and ends with the last phase slip at \(t \approx 14.247\). During this period, the phase difference \(\Phi\) between the edges of the segment along the \(x\)-axis decreases from its initial value of \(2\pi \cdot 16\) to a final value of \(2\pi\). This stage corresponds to the most significant changes in the Lyapunov functional, as shown in Fig.~\ref{fig:FGL}.

{\bf (iv)} After the phase slip generation stage concludes, the dynamics enter a final stage characterized by a slow relaxation to a stable stationary solution, described by expressions \eqref{eq:psiGL}, with a single selected wavenumber \(k\). In the case under discussion, it equals $\Delta k$.

\begin{figure}[h!]
  \centering
  \includegraphics[width=.7\columnwidth]{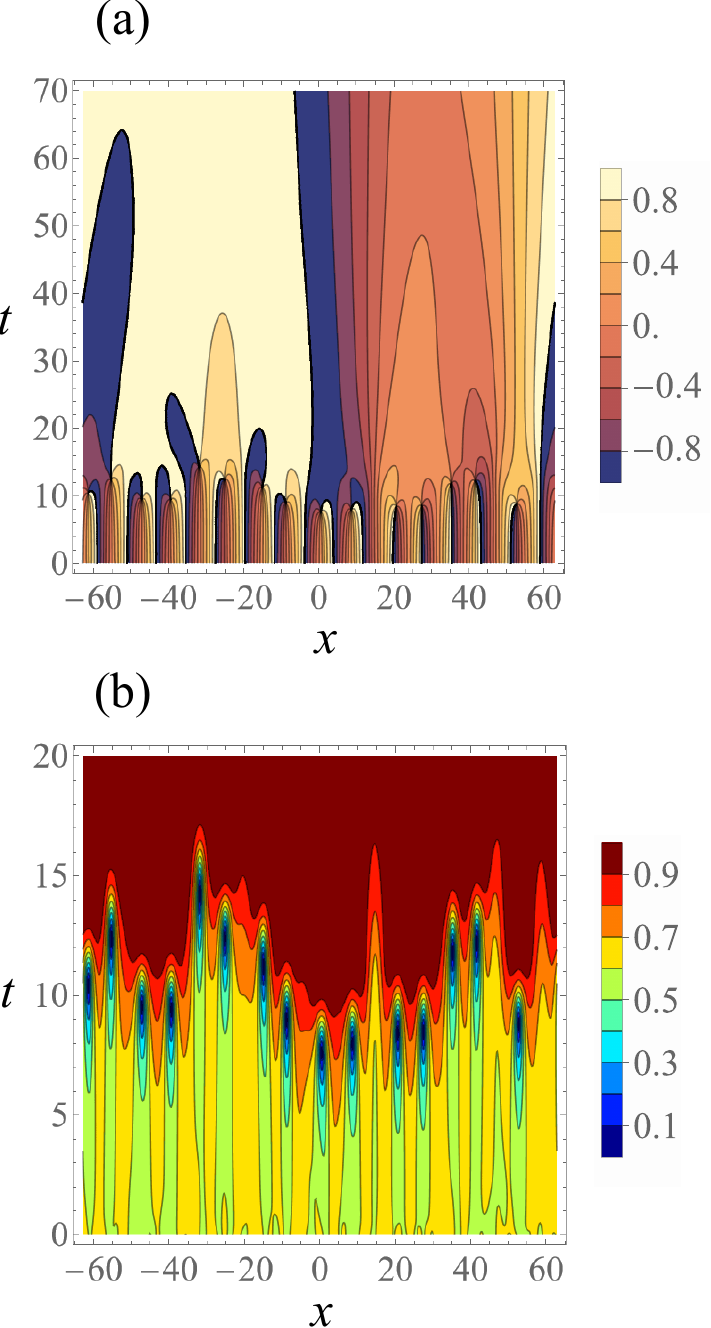}
  \caption{Contour plots of (a) ${\rm Arg}[\psi(x,t)]/\pi$ and (b) $|\psi(x,t)]|$. The phase is reduced to the domain $[-\pi,\pi]$. Vanishing of $\psi(x,t)$ at certain values of $x$ and $t$ results in phase slips. The phase slip generation occurs at \mbox{$7.606<t<14.247$}. It corresponds to sharp phase dynamics. Note that the variations of $|\psi|$ are, practically, over after a short relaxation process associated with the last phase slip, which happens at \mbox{$x \approx -31.838$}, \mbox{$t\approx 14.247$}. In contrast, at that moment, the phase only changes its dynamics --- the sharp phase variations at \mbox{$t<14.247$} are replaced by slow relaxation to the stationary state at \mbox{$t>14.247$}, cf. panels (a) and (b).
  } \label{fig:phase_abs_Contours}
\end{figure}

Note however, that while the modulus of this final selected wavenumber always is close to the expression \mbox{$|k_0|-|q_{\rm max}|$,} where $q_{\rm max}$ maximizes the Eckhaus growth rate $\sigma(k_0,q)$, it is not unique for a given $k_0$ and depends on the set of the initial perturbations. In the simulations, various runs of the numerical code differ only by the values of the random phases in the set of initial perturbations may produce close but different values of the final wavenumber. 

Note also that the $|\psi|$ profile has practically relaxed to its asymptotical form by $t=20$. In contrast, the phase profile had only smoothed down its sharp modulations by this time. Look at the corresponding plots in Fig.~\ref{fig:t20}. 
Contour plots of the phase and modulus of $\psi(x,t)$ presented in Fig.~\ref{fig:phase_abs_Contours} illustrate these results, making them visual. 

\begin{figure}[h!]
  \includegraphics[width=\columnwidth]{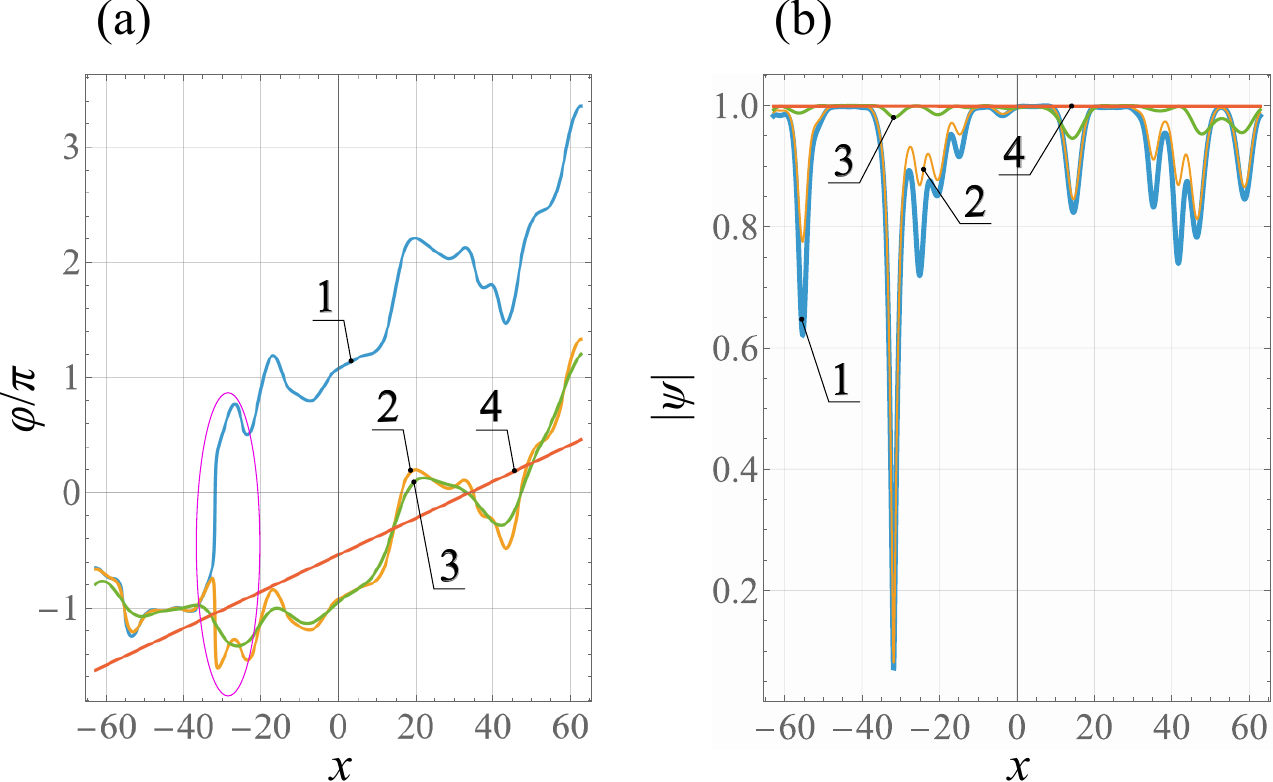}
   \caption{Profiles of the phase (a) and modulus (b) of $\psi$ at various characteristic moments. Numbers 1 -- 4 indicate the moments \mbox{$t=14$}; \mbox{$t=14.5$}; $t=20$; and the asymptotical state at $t\to \infty$, respectively. See the text for details.}\label{fig:ps14}
\end{figure}

In Fig.~\ref{fig:ps14}, we illustrate the phase slip process more demonstratively by displaying the profiles of the modulus of $\psi$ and its phase (in continuous, unreduced form) at several key moments. These moments include: 
\begin{itemize}
  \item $t=14$ — just before the last phase slip occuring at $t \approx 14.247$; 
  \item $t=14.5$ — just after the last phase slip; 
  \item $t=20$ — after the phase slip stage has concluded and the local perturbations caused by the last phase slip have relaxed; 
  \item the asymptotic state corresponding to the stationary solution with $k=\Delta k$. 
\end{itemize}

The vicinity of the phase slip is marked with an oval. It is evident that the phase slip results in the $2\pi$ shift of the entire phase profile to the right of the phase slip point at $x \approx 30.994$, while the portion to the left of this point does not change.

The only remaining issue to discuss is the dynamics of the Fourier coefficients $|\psi_k(t)|$ for different values of $k$. For $k$'s corresponding to the stable modes in the Eckhaus spectrum, the dynamics are straightforward --- the modes decay over time. 

The dynamics of the mode with the wavenumber corresponding to the final selection is also simple. Initially, its amplitude grows exponentially at a rate close to that following from Eq.~\eqref{eq:GL_sigma(k,p)}. Then, during the phase slip generation stage, the growth rate transforms into the one describing the final relaxation to the pattern with the single selected wavenumber; see curve (1) in Fig.~\ref{fig:psi_0.1_1_1.55}. 

Modes with wavenumbers close to the finally selected one display similar behavior. The only difference is that instead of becoming saturated, their amplitudes eventually decay and vanish asymptotically. This decay process is very slow --- the slower, the closer the mode's wavenumber to that of the asymptotic spatially periodic stationary pattern. For instance, in the case under discussion, the finally selected wavenumber corresponds to $k=\Delta k = 0.05$. Consequently, for the mode with $k=2\Delta k$, the decay does not begin until approximately $t \approx 62.284$.

The dynamics associated with other unstable modes of the linear Eckhaus spectrum are less straightforward. Initially, their evolution is influenced by two competing factors: linear Eckhaus instability and the nonlinear stabilization due to coupling with stable modes. In simulations conducted under the specified conditions, for the wavenumbers that are not too close to the eventually selected one, initially the nonlinear stabilization tends to dominate, leading to a decay in the amplitudes of these modes. 

However, during the period designated above as stage (i), the stable modes also experience rapid decay, decreasing the nonlinear stabilization. Meanwhile, the linear Eckhaus instability remains practically unchanged, as all modes still stay within the range where the linear Eckhaus stability analysis is applicable. Consequently, by the end of stage (i), nonlinear decay is overtaken by exponential growth, governed by the growth rate described by Eq.~\eqref{eq:GL_sigma(k,p)}. This growth continues until the onset of the phase slip stage, at which point wavenumber selection begins and the spectrum contracts to the final selected wavenumber. As a result, all other modes start to decay. Curves (2) and (3) in Fig.~\ref{fig:psi_0.1_1_1.55} illustrate this typical behavior.

\section{Conclusions}

The above analysis has provided a comprehensive overview of the evolution of the Eckhaus instability, which is triggered by small-amplitude, broad-spectrum initial perturbations. This study tracks the process from its linear stage to the final wavenumber selection. It identifies, categorizes, and clarifies four distinct stages of evolution, linking their characteristics to the behavior of the order parameter in both coordinate and Fourier spaces and to the Lyapunov functional. The findings appear to be general and offer fresh insights into this longstanding and significant issue.

\section{Acknowledgements}

The author would like to express gratitude to Boris Y. Rubinstein for his valuable assistance with programming.

  \bibliographystyle{elsarticle-num} 
  \bibliography{Eckhaus}



%
%
%
\end{document}